\documentclass[aps,prl,preprint,superscriptaddress,showpacs]{revtex4}
\usepackage{graphicx}

\newlength{\intwidth}

\def\lapprox{\,\raise0.4ex\hbox{$<$}\kern-0.8em\lower0.7ex\hbox{$\sim$}\,}
\def\gapprox{\,\raise0.4ex\hbox{$>$}\kern-0.8em\lower0.7ex\hbox{$\sim$}\,}
\def\lg{\,\raise0.5ex\hbox{\footnotesize $<$}\kern-0.8em\lower0.5ex\hbox{\footnotesize  $>$}\,}
\def\gl{\,\raise0.5ex\hbox{\footnotesize $>$}\kern-0.8em\lower0.5ex\hbox{\footnotesize  $<$}\,}

\def\be{\begin{equation}}
\def\ee{\end{equation}}
\def\ba{\begin{eqnarray}}
\def\ea{\end{eqnarray}}

\def\bc{\begin{center}}
\def\ec{\end{center}}

\begin{document}

\title{Extra Spin-Wave mode in Quantum Hall systems. Beyond the Skyrmion Limit.}

\author{I.K.~Drozdov}
\author{L.V.~Kulik}
\author{A.S.~Zhuravlev}
\author{V.E.~Kirpichev}
\affiliation{Institute of Solid State Physics, RAS, Chernogolovka, 142432 Russia}
\author{I.V.~Kukushkin}
\affiliation{Institute of Solid State Physics, RAS, Chernogolovka, 142432 Russia}
\affiliation{Max-Planck-Institut f\"ur Festk\"orperforschung, Heisenbergstr. 1, 70569 Stuttgart,Germany}
\author{S.~Schmult}
\author{W.~Dietsche}
\affiliation{Max-Planck-Institut f\"ur Festk\"orperforschung, Heisenbergstr. 1, 70569 Stuttgart,Germany}
\date{\today}

\begin{abstract}
We report on the observation of a new spin mode in a quantum Hall system in the vicinity of odd electron filling factors under experimental conditions excluding the possibility of Skyrmion excitations. The new mode having presumably zero energy at odd filling factors emerges at small deviations from odd filling factors and couples to the spin-exciton. The existence of an extra spin mode assumes a nontrivial magnetic order at partial fillings of Landau levels surrounding quantum Hall ferromagnets other then the Skyrmion crystal.
\end{abstract}


\maketitle Phenomena associated with spontaneous magnetic order in two dimensions continue to be a rich source of new physical theories. Among them, the concept of the quantum Hall ferromagnet is probably the most elaborated. It is well established that the ground state of a two-dimensional electron system at odd integer filling factors is an itinerant ferromagnet \cite{PinczDasSarm}. The quantum Hall ferromagnet can support a single Goldstone or (at finite Zeeman coupling) pseudo-Goldstone mode, which is a spin-exciton composed of an excited electron in the empty spin branch of the Landau level and a hole in the filled branch with an opposite spin \cite{bychkov,kallin84}. According to the Larmor theorem, a zero momentum spin-exciton has the energy equal to the bare Zeeman gap, whereas its energy at the infinite momentum is defined by the exchange interaction. The exchange energy for experimentally accessible electron systems in AlGaAs/GaAs quantum wells (QWs) exceeds the Zeeman energy by some 2 orders of magnitude. The long ferromagnetic order nevertheless persists only at low temperatures below the Zeeman gap. At larger temperatures, proliferation of low energy spin-excitons breaks the ferromagnet apart into clusters. Lateral confinement of those clusters pushes the energies of spin-excitons up, thus preserving the ferromagnetic order locally \cite{nu1our}.

 This physically intuitive picture does not hold for non-integer filling factors. The ground state at $\nu$ slightly away from unity contains a finite density of Skyrmions---spin-texture excitations with the spin projection along the magnetic field distorted smoothly in a vortex-like configuration \cite{sondhi}. The Skyrmion core radius $R$ and the effective number of spin reversals $K$ are characteristic parameters of the inhomogeneity of spin rotation in real space. Their values are a result of competition between the loss of Zeeman energy, when a large number of spins are tilted off the magnetic field axis, and the gain of lowering the exchange energy between neighboring spins. Skyrmions are relevant to the physics of electron spins if $K\gg1$. For experimentally accessible cases $K\approx 3$ a quantum theory of "spin-texture quasiparticles" (small radius Skyrmions), which are a natural generalization of classical Skyrmion states, has been developed \cite{Abolfath}. The formation of small radius Skyrmions is confirmed by nuclear magnetic resonance and optical absorption experiments \cite{Khandelwal,Aifer}. More recent works exploiting the trion dichroism strengthened confidence in the validity of the small Skyrmion picture at noninteger filling factors \cite{Groshaus}.

The Skyrmion-Skyrmion interaction may lead to the formation of a Skyrme crystal---a periodic lattice formed of Skyrmion and anti-Skyrmion excitations \cite{Cote}. The Skyrme crystal breaks the rotational symmetry of an individual Skyrmion around the magnetic field axis. As a result, an extra Goldstone mode related to a global spin rotation in the $XY$ plane evolves. Having a unit spin, the new mode remains gapless in the presence of a Zeeman field and is responsible for rapid nuclear spin relaxation \cite{Tycko,Smet}. In addition, the Skyrme crystal manifests itself through lowering of the spin-exciton energy at partial fillings of the zero spin Landau level \cite{Gallais}.

Despite the growing experimental proof for the relevance of Skyrmions to physics of non-integer filling factors, no direct evidence of the extra spin Goldstone mode has ever been presented. Besides, it is unclear what happens with Skyrmions when their number of reversed spins reaches the $K=0$ limit. With a large portion of confidence, one can affirm that the Skyrmion with $K=0$ is reduced to a single hole in a partially filled Landau level (Laughlin quasiparticle). Therefore, this limiting case presents a tremendous experimental significance as the first step towards understanding the physics of the more tangled cases $K\geq1$.

In the present Letter we show that an extra spin mode exists at filling factors slightly away from unity even if $K=0$, i.e., beyond the realm of Skyrmion model. It is not truly a Goldstone mode since its energy reaches finite values at zero momentum. The new spin mode couples to the spin-exciton from either side of $\nu=1$ with the coupling strength growing with momentum. Most surprising is that it exists at odd filling factors other than $\nu=1$, where Skyrmions, as was proved theoretically, could not form even without Zeeman field \cite{Fertig}. The existence of an extra spin mode signals the formation of a novel type of magnetic order at partial fillings of Landau levels, whereas the finite energy of this mode suggests that this magnetic order persists only locally ({\it spin texture liquid}).

To establish experimental conditions for the $K=0$ limit, we applied a large magnetic field to a sample with a low electron density. The magnetic field was oriented at small angle to the sample surface. The large in-plane component of the magnetic field produces a substantial Zeeman field for electrons, whereas the low electron density guarantees a sufficiently small Coulomb energy. Narrow GaAs/AlGaAs quantum wells of 20 nm width hosting a high-mobility two-dimensional electron system were used (the mobility in the dark was $5\times 10^6$ cm$^2$/Vs). The samples were chosen to maximize energy gaps separating the size-quantized electron subbands and thus to mitigate the subband mixing induced by the in-plane magnetic field. The electron density $n_s$ in the QWs was tuned to any desired value between $0.5\times 10^{10}$ and $6\times 10^{10}$ cm$^{-2}$ via the optodepletion effect and was measured by means of {\it in situ} photoluminescence and inelastic light scattering \cite{Kulik}. The in-plane magnetic field was varied continuously at the fixed total magnetic field with the help of a rotational stage. We reached precision in angle between the sample surface and the magnetic field better than $0.1^\circ$. The experiment ran as follows: having fixed the total magnetic field and the sample tilt angle, we varied the intensity of the photodepleting laser to adjust the filling factor.

To access low energy spin excitations. the inelastic light scattering (ILS) technique was employed. The ILS experiment was performed at the temperature of 0.3 K in the magnetic field of 14.5~T. the ILS process transferred a momentum to excitations of the electron system. Because of the small QW electron density, the ILS momentum reached values up to half of the reciprocal interparticle distance at which spin excitations manifest their Coulomb nature. ILS spectra were obtained using a laser tunable above the fundamental band gap of GaAs. The scattered light was dispersed by a triple spectrograph and recorded with a charge-coupled device camera. The overall spectral resolution of the detection system was 0.04 meV.

\begin{figure}[htb!]
\includegraphics[width=7.2cm]{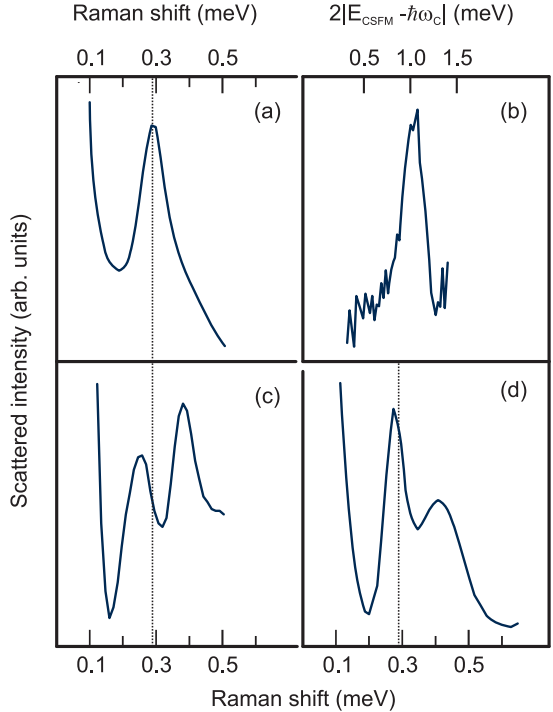}
\caption{\label{fig1} ILS spectrum of spin exciton at $\nu=1.9$ which measures directly the bare Zeeman energy (a). ILS spectrum of the cyclotron spin-flip mode (CSFM) at $\nu=1$ vs doubled difference between the CSFM energy and the cyclotron energy which gives approximately the exchange energy per particle (b). ILS spectra of two spin-wave modes at electron filling factors above ($\nu=1.35$) (c) and below ($\nu=0.75$)(d) the quantum Hall ferromagnet $\nu=1$. The dotted lines show the bare Zeeman energy.}
\end{figure}

To free the experimental results from any preliminary assumptions concerning the ratio between the exchange and the Zeeman energies, we utilize the technique developed in Ref.[\onlinecite{nu1our}] to measure the exchange energy directly. This is absolutely necessary as the exchange interaction strength declines significantly from its value obtained within the standard theoretical approach at the electron densities below $3\times 10^{10}$ cm$^{-2}$ most probably due to the presence of barrier impurities and the formation of $D^-$ complexes \cite{nu1our}. For example, the exchange energy at $n_s=2\times 10^{10}$ cm$^{-2}$ is about 1 meV which is nearly 3 times less than the theoretically predicted value (Fig.~1). The Zeeman coupling is also controlled directly through the spin-wave energy measured at small ILS momenta far from odd electron filling factors, i.e., under the experimental conditions which guarantee negligible Coulomb corrections to the Larmor mode energy (Fig.~1). From the measured quantities, the ratio of the Zeeman and the exchange energies is calculated. It rises almost to $1/3$ at $n_s=2\times 10^{10}$ cm$^{-2}$ which is 5 times larger than the theoretically predicted ratio for the $K=0$ limit \cite{Abolfath}. Relying on the existing theoretical background, we should exclude the possibility of forming small texture quasiparticles as well as a Skyrme crystal under our experimental conditions.

It is tempting to think that at so small, if compared to the Zeeman gap, exchange energies, the ground state should be a collinear ferromagnet with a finite number of defects---quasiholes. Nevertheless, the anomaly in the spectrum of spin excitations evidences that the collinear ferromagnet concept fails, Fig.~1. Our major experimental finding is the observation of the spin-exciton spectral component splitting in two lines at electron filling factors slightly away from $\nu=1$. One line lies below whereas another one lies above the Zeeman energy (Fig.~1). Besides, a similar splitting is detected at electron filling factors slightly away from $\nu=3$ (Figure~2). The intensities and energies of both lines demonstrate a clear anticrossing behavior. The energy of one line tends to the bare Zeeman energy at filling factors far away from $\nu=1(3)$ and approaches zero in close proximity to the odd filling factors. The other line corresponds to the low momentum spin-exciton at $\nu=1(3)$ and rises at small deviations from the odd filling factors (Fig.~3).

\begin{figure}[htb!]
\includegraphics[width=4.8cm]{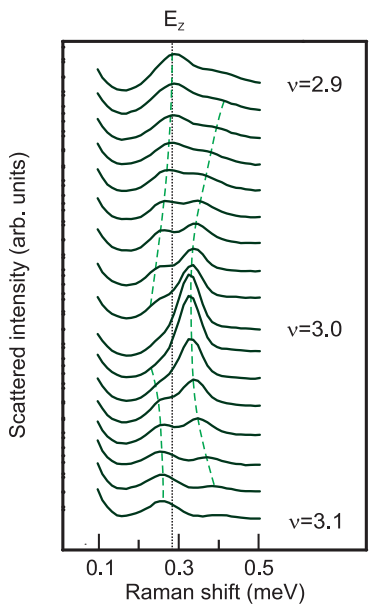}
\caption{\label{fig2} ILS spectra around the quantum Hall ferromagnet $\nu=3$ with a step in the electron filling factor of about 0.0125. The dotted line shows the bare Zeeman energy. The dashed line is a guide to the eye.}
\end{figure}

The following tests validate that both spin-wave modes have a Coulomb nature, as well as that they are truly eigenmodes of the two-dimensional electron system at non-integer filling factors. First, the coupling strength between two modes changes with the electron density at fixed ILS momentum (Fig.~3). This implies the collective Coloumb nature for both modes. Second, the coupling strength at the fixed electron density varies with the ILS momentum (Fig.~4). The last observation provides a clear evidence that both modes are excited states with a definite momentum; i.e., they are not associated with localized states (e.g. $D^-$ centers \cite{nu1our}). Finally, the Larmor theorem is obviously satisfied. It is apparent from Figs.~3 and 4, that the higher energy spin-wave mode is transformed into the spin-exciton pseudo Goldstone mode of the quantum Hall ferromagnet state $\nu=1(3)$. Its energy scales with the square of the ILS momentum at $\nu=1(3)$ and reaches the bare Zeeman gap at zero momentum (Fig.~3).

\begin{figure}[htb!]
\includegraphics[width=7.2cm]{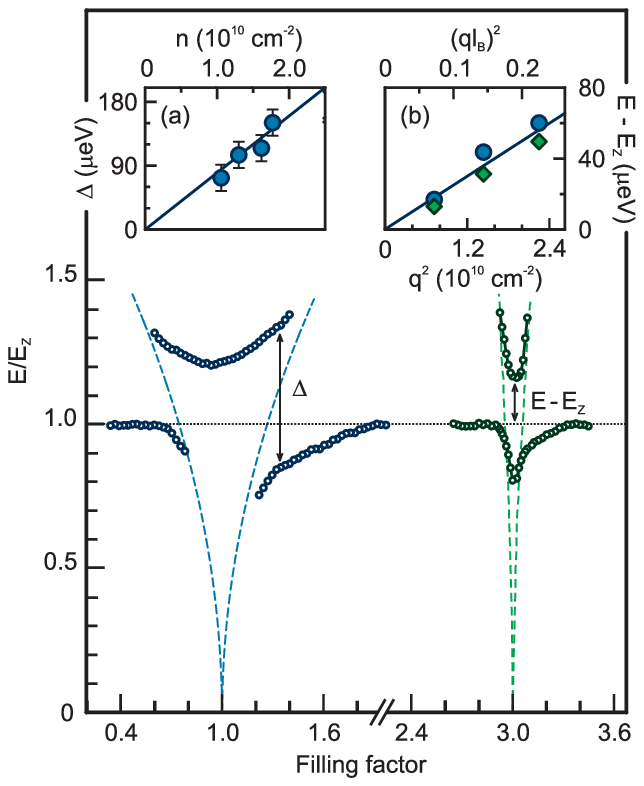}
\caption{\label{fig3} The energies of ILS lines vs electron filling factors around the quantum Hall ferromagnets $\nu=1$ and $\nu=3$ (open circles). The dotted line shows the bare Zeeman energy. The dashed lines are $\sqrt{|1-\nu|}$ and $\sqrt{|3-\nu|}$ dependencies illustrating possible trends for the energy of the extra spin-wave mode. The insets show the dependence of the minimal splitting of the two spin-wave modes on the electron density measured at $\nu>1$ (a), and the experimental dispersion for the spin-exciton in the quantum Hall ferromagnets $\nu=1$ (circles) and $\nu=3$ (diamonds) on the square of ILS momentum (b). The solid lines are guides to the eye.}
\end{figure}

As a collinear ferromagnet does not support two independent spin-wave modes, one is forced to accept an extra non-collinear magnetic order in close proximity to the odd electron filling factors $\nu=1(3)$ at so small ratios of the exchange and the Zeeman energies as they are in our experiment. The nature of the new magnetic order is yet unclear; however, some assumptions concerning the ordered state can be put forward. Note that the ratio between Zeeman and exchange energies is kept nearly constant while the electron filling factor is varied around $\nu=1(3)$ \cite{exchange}. However, namely, this ratio determines the spin structure of the ground state spin-texture defects whatever they could be. Moving away from the odd filling factors, one does not change the spin structure of those defects, only the inter-defect separation scales as $\sqrt{|1-\nu|}$ ($\sqrt{|3-\nu|}$). Assuming that the energy of the extra spin-wave mode increases continuously with $|1-\nu|(|3-\nu|)$ \cite{ins} and that it is associated with Coulomb interaction between the defects, we come to an intuitive square-root dependence for the mode energy as a function of the filling factor. Hopefully a properly developed theory will explain the experimental findings in more appropriate terms.

\begin{figure}[htb!]
\includegraphics[width=7.2cm]{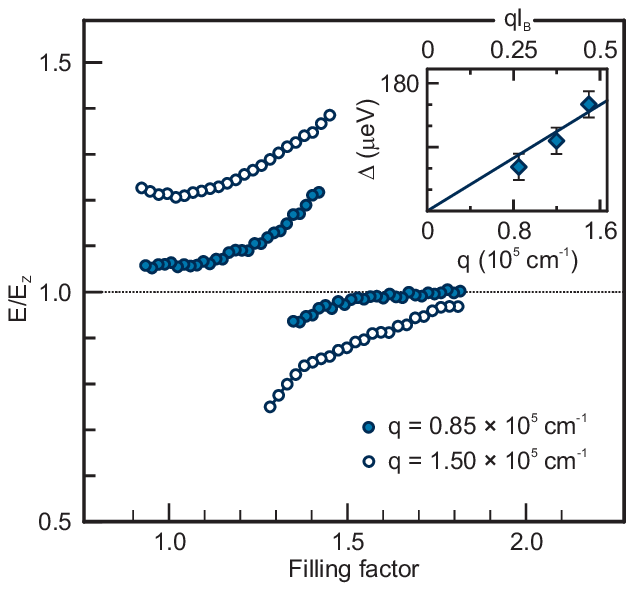}
\caption{\label{fig4} Energies of ILS lines vs the electron filling factor in the vicinity of the quantum Hall ferromagnet $\nu=1$ at two different ILS momenta. The dotted line shows the bare Zeeman energy. The inset shows the experimental dispersion for the minimal splitting between two spin-wave modes vs ILS momentum. The solid line is a guide to the eye.}
\end{figure}

To conclude, we outline briefly the physical importance of the reported results. The Skyrmion concept is well elaborated in the specific regime $K\gg1$. In high-mobility quantum well GaAs/AlGaAs heterostructures---systems where electron-electron interaction plays the major role for the formation of the ground state---the effective number of reversed spins for a Skyrmion generally lies between 3 and 4. Some new experiments demonstrate a much larger Skyrmion spin up to 12 \cite{Plochocka}. This contradicts variational simulations and is in fact questioning the validity of the Skyrmion model. To obtain a consistent view of the spin ground state in the whole range of ratios between the Zeeman and the Coulomb energies one has to start from a well-defined limit $K=0$ and move on to more intricate cases. Presumably, the ground state with $K=0$ contains single holes as spin defects. Whether those holes form a crystal lattice or not is an open question, but in any case, the ground state should be a collinear ferromagnet supporting a single pseudo-Goldstone spin-wave mode that should obey the Larmor theorem. Already, this simplified system demonstrates surprisingly rich physics, which could hardly be expected aforehand. A new spin-wave mode with unusual properties is observed. It is not truly a Goldstone mode; however, its energy can fall well below the Zeeman gap, and it scales continuously with the electron filling factor. This assumes that the ground state of a two-dimensional system at non-integer filling factors is a short-range non-collinear ferromagnet. Thus, we discover a novel broken symmetry ground state other than the Skyrmion crystal which supports an extra spin-wave mode. Since the energy of this mode falls below the Zeeman gap, it should be related to spin fluctuations in the $XY$ plane.

We thank S.~Dickmann and A.~Pinczuk for many helpful discussions. This work was supported by the BMBF, the DFG, and the Russian Fund of Basic Research.


\begin{thebibliography}{99}
\bibitem{PinczDasSarm} {\it Perspectives in Quantum Hall Effects},
edited by S.Das Sarma and A.Pinczuk (Wiley, New York, 1997).

\bibitem{bychkov}
Yu. A. Bychkov, S. V. Iordanskii, and G. M. Eliashberg, JETP Lett.~{\bf
33}, 143 (1981).

\bibitem{kallin84}
C. Kallin, and B. I. Halperin, Phys. Rev. B {\bf 30}, 5655 (1984).

\bibitem{nu1our}
A.B.~Van'kov, L.V.~Kulik, I.V.~Kukushkin, V.E.~Kirpichev, S.~Dickmann, V.M.~Zhilin,
J.H.~Smet, K.~von~Klitzing, and W.~Wegscheider, Phys.~Rev.~Lett. {\bf 97}, 246801 (2006);
A.S.~Zhuravlev, A.B.~Van'kov, L.V.~Kulik, I.V.~Kukushkin, V.E.~Kirpichev, J.H.~Smet,
K.~von~Klitzing, V.~Umansky, and W.~Wegscheider, Phys.~Rev.~B {\bf 77}, 155404 (2008).

\bibitem{sondhi}
S.~L.~Sondhi, A.~Karlhede, S.~A.~Kivelson, and E.~H.~Rezayi, Phys.~Rev.~B {\bf 47}, 16419 (1993).

\bibitem{Abolfath}
M. Abolfath, J. J. Palacios, H. A. Fertig, S. M. Girvin and A. H. MacDonald, Phys. Rev. B {\bf 56}, 6795 (1997).

\bibitem{Khandelwal}
S.~E.~Barrett, G.~Dabbagh, L.~N.~Pfeiffer, K.~W.~West, and R.~Tycko, Phys. Rev. Lett.~{\bf 74}, 5112 (1995);
P. Khandelwal, A. E. Dementyev, N. N. Kuzma, S. E. Barrett, L. N. Pfeiffer, and K.W. West, Phys.~Rev.~Lett. {\bf 86}, 5353 (2001).

\bibitem{Aifer}
E. H. Aifer, B. B. Goldberg, D. A. Broido, Phys.~Rev.~Lett. {\bf 76}, 680 (1996).

\bibitem{Groshaus}
J. G. Groshaus, V. Umansky, H. Shtrikman, Y. Levinson, and I. Bar-Joseph, Phys.~Rev.~Lett. {\bf 93}, 096802 (2004);
J. G. Groshaus, P. Plochocka-Polack, M. Rappaport, V. Umansky, I. Bar-Joseph, B. S. Dennis, L. N. Pfeiffer, K. W. West, Y. Gallais, and A. Pinczuk,
Phys.~Rev.~Lett. {\bf 98}, 156803 (2007).

\bibitem{Cote}
R. Cote, A. H. MacDonald, Luis Brey, H. A. Fertig, S. M. Girvin, and H. T. C. Stoof, Phys.~Rev.~Lett. {\bf 78}, 4825 (1997).

\bibitem{Tycko}
R. Tycko, S. E. Barrett, G. Dabbagh, L. N. Pfeiffer, and K. W. West, Science {\bf 268}, 1460 (1995).

\bibitem{Smet}
J. H. Smet, R. A. Deutschmann, F. Ertl, W. Wegscheider, G. Abstreiter, and K. von Klitzing, Nature {\bf 415}, 281 (2002).

\bibitem{Gallais}
Y. Gallais, J. Yan, A. Pinczuk, L. N. Pfeiffer, and K. W. West, Phys.~Rev.~Lett. {\bf 100}, 086806 (2008).

\bibitem{Fertig}
H. A. Fertig, L. Brey, R. Cote, A. H. MacDonald, A. Karlhede, S. L. Sondhi, Phys.~Rev.~B {\bf 55}, 10671 (1997).

\bibitem{Kulik}
L. V. Kulik, S. Dickmann, I. K. Drozdov, A. S. Zhuravlev, V. E. Kirpichev, I.. V. Kukushkin, S. Schmult, and W. Dietsche, Phys. Rev. B {\bf 79}, 121310(R) (2009).

\bibitem{exchange}
Zeeman energy is kept constant while the ground state exchange energy varies slightly in the close vicinity of $\nu=1$ as $\sqrt{\nu}$.

\bibitem{ins}
We do not see any physical reasons allowing the spin-wave mode energy to jump discontinuously with the infinitely small change of the electron filling factor off the $\nu=1(3)$ ferromagnets (see the illustration to Figure~3).

\bibitem{Plochocka}
P. Plochocka, J. M. Schneider, D.K. Maude, M. Potemski, M. Rappaport, V. Umansky, and I. Bar-Joseph, J. G. Groshaus, Y. Gallais, and A. Pinczuk, Phys.~Rev.~Lett. {\bf 102}, 126806 (2009).

\end{thebibliography}
\end{document}